# Strain-mediated phase crossover in Ruddlesden–Popper nickelates

Ting Cui,[1,2] Songhee Choi,[1] Ting Lin,[1,2] Chen Liu,[3] Gang Wang,[1,2] Ningning Wang,[1,2] Shengru Chen,[1,2] Haitao Hong,[1,2] Dongke Rong,[1,2] Qianying Wang,[1,2] Qiao Jin,[1] Jia-Ou Wang,[3] Lin Gu,[4] Chen Ge,[1] Can Wang,[1,2,5] Jin Guang Cheng,[1] Qinghua Zhang,[1,2,5] Liang Si,[6,7,*] Kui-juan Jin,[1,2,5,*] and Er-Jia Guo[1,2,*]

[1] Beijing National Laboratory for Condensed Matter Physics and Institute of Physics, Chinese Academy of Sciences, Beijing 100190, China

[2] Department of Physics & Center of Materials Science and Optoelectronics Engineering, University of Chinese Academy of Sciences, Beijing 100049, China

[3] Institute of High Energy Physics, Chinese Academy of Sciences, Beijing 100049, China

[4] National Center for Electron Microscopy in Beijing and School of Materials Science and Engineering, Tsinghua University, Beijing 100084, China

[5] Songshan Lake Materials Laboratory, Dongguan, Guangdong 523808, China

[6] School of Physics, Northwest University, Xi'an 710127, China

[7] Institut für Festkörperphysik, TU Wien, 1040 Vienna, Austria

*Correspondence and requests for materials should be addressed to Q.H.Z. and E.J.G. (emails: siliang@nwu.edu.cn, kjjin@iphy.ac.cn and ejguo@iphy.ac.cn)

**Abstract**

Recent progress on the signatures of pressure-induced high-temperature superconductivity in Ruddlesden–Popper (RP) nickelates ($La_{n+1}Ni_nO_{3n+1}$) has attracted growing interest in both theoretical calculations and experimental efforts. The fabrication of high-quality single-crystalline RP nickelate thin films is critical for possible reducing the superconducting transition pressure and advancing applications in microelectronics in the future. In this study, we report the observations of an active phase transition in RP nickelate films induced by misfit strain. We found that RP nickelate films favor the perovskite structure ($n = \infty$) under tensile strains, while compressive strains stabilize the $La_3Ni_2O_7$ ($n = 2$) phase. The selection of distinct phases is governed by the strain dependent formation energy and electronic configuration. In compressively strained $La_3Ni_2O_7$, we experimentally determined the $e_g$ splitting energy is ~0.2 eV and electrons prefer to occupy in-plane orbitals. First-principles calculations unveil a robust coupling between strain effects and the valence state of Ni ions in RP nickelates, suggesting a dual driving force for the inevitable phase co-existence transition in RP nickelates. Our work underscores the sensitivity of RP nickelate formation to epitaxial strain, presenting a significant challenge in fabricating pure-phase RP nickelate films. Therefore, special attention to stacking defects and grain boundaries between different RP phases is essential when discussing the pressure-induced superconductivity in RP nickelates.

**Main text**

The discovery of infinite-layered nickelates in both the parent compound and with variable dopants has spurred research in the quest for new unconventional superconductors analogous to cuprates [1-6]. The $d^9$ configuration of $Ni^{1+}$ in infinite-layered nickelates shares many similarities with that of cuprates, making it a focal point for investigating the underlying mechanisms of high-temperature (HT) superconductivity [7,8]. Up to now, the highest superconducting transition temperature ($T_C$) achieved in infinite-layered nickelate films was approximately 20 K at ambient pressure, and it increased to slightly above 30 K under high pressure [9]. In addition to infinite-layered nickelates, the Ruddlesden–Popper (RP) phase nickelates ($La_{n+1}Ni_nO_{3n+1}$) present themselves as HT superconducting candidates. The large electronic bandwidth and strong correlation between Ni $3d$ and O $2p$ bands hold promise for enhancing $T_C$. Recently, Sun *et al*. reported the signatures of HT superconductivity with $T_C$ exceeding 80 K in $La_3Ni_2O_7$ (RP phase with $n = 2$) under a high pressure of 14 GPa [10]. Suddenly, much attention has shifted to these new materials, not only due to their $T_C$ surpassing the boiling temperature of liquid nitrogen but also because of their intriguing electronic configuration [11-19], which is dramatically different compared to the well-known cuprates [20-24].

Presently, one of the research priorities involves the fabrication of high-quality single-crystalline Ruddlesden–Popper (RP) nickelate thin films with a controllable number of layers and precise oxygen stoichiometry. This is essential for various aspects, including quantum confinement at the heterointerfaces and gaining a comprehensive understanding of the underlying mechanisms of pressure-induced HT superconductivity in nickelates. To begin with, the orientation of substrates plays a pivotal role in regulating the growth direction of the thin films [25,26]. This interplay between superconductivity and other emerging quantum states with broken symmetry and anisotropy can be thoroughly investigated. Secondly, misfit strain introduces the in-plane biaxial structural modifications to the unit cells [27-29]. Applying biaxial compressive strain to thin films is analogous to subjecting them to uniaxial pressure. The bonding geometry directly influences the strength of orbital hybridization, thus it is anticipated that the superconducting transition pressure in RP nickelates will be reduced, hopefully down to ambient pressure. Thirdly, emergent interface phenomena, including orbital reconstruction, charge transfer, and octahedral fine tuning across heterointerfaces, may alter the ground states of RP nickelates [30-32], leading to an effective pathway to enhance the electronic correlations and to modulate the orbital occupations through interfacial coupling. Consequently, the fabrication of high-quality thin films will significantly enrich the phase diagram of RP nickelate superconductors, advancing their potential applications in microelectronic devices in the future.

Earlier studies have shown that the phase formation in oxide thin films through pulsed laser deposition (PLD) is significantly influenced by growth conditions, including substrate temperature, growth rate, laser fluence, and oxygen partial pressure [33,34]. In addition to these deposition parameters, misfit strain from substrates is known as a crucial factor for tuning oxygen stoichiometry [35-37]. Creating/compensating oxygen vacancies equals to the electron/hole doping into the equivalent state; thus the electronic configuration of transition metal ions could be largely influenced. In the particular case, the crystal structure of RP nickelates can be tuned from low order (small $n$) to high order (large $n$) phases with the nominal valence state of Ni ion shifts from $Ni^{2+}$ ($n = 1$) to $Ni^{3+}$ ($n = \infty$) and $d$ orbital occupation from $d^8$ ($n = 1$) to $d^7$ ($n = \infty$). A comprehensive understanding of the direct relationship between strain and the formation energy of distinct phases is imperative for attaining the desired RP nickelates. In this study, we demonstrate the stabilization of different RP-phase nickelate thin films by altering misfit strain under identical deposition conditions. The choice of phases is dictated by the formation energy, which is contingent on the epitaxial strain. The intricate structural and electronic states observed in RP nickelates, potentially contributing to the observed superconductivity under high pressure, are rooted in the direct connection between elastic strain and phase formation.

**Results and discussions**

In the realm of RP nickelates, the configuration comprises corner-shared $NiO_6$ octahedra with a variable number of $n$ layers, separated by the LaO-LaO rock-salt structures. These naturally disrupt the chemical bonding between adjacent octahedra. By adjusting the $n$ value, the crystalline structure can transit from a three-dimensional perovskite arrangement ($n = \infty$,

Figure 1a) to a layered structure ($n = 2$, Figure 1b), and further to a two-dimensional structure ($n = 1$). Figure 1c shows X-ray diffraction (XRD) $\theta$-$2\theta$ curves of RP nickelate thin films around the substrates' peaks. The in-plane misfit strain ($\varepsilon$) of RP nickelate films can be tuned from 3.44% (tensile strain) to –4.12% (compressive strain). Under compressive strain conditions, the RP nickelates exhibit a typical $La_3Ni_2O_7$ ($n = 2$) structure, identical to the composition of the ablation target. Surprisingly, when subjected to tensile strain, the RP nickelate films exhibit a perovskite structure ($n = \infty$). The distinct thickness fringes around the film peaks indicate excellent crystalline quality. Wide-range XRD $\theta$-$2\theta$ curves of RP nickelate films on $DyScO_3$ (DSO) and $LaAlO_3$ (LAO) substrates confirm different crystal structures under various strain states (Figure S1). Figures 1d and S2 illustrate the XRD reciprocal space maps (RSMs) of RP nickelate films. Within the moderate strain region, RP nickelate films maintain coherent strain alignment with the substrates. However, films grown on $SrLaAlO_4$ (SLAO) and $YAlO_3$ (YAO) substrates slightly relax the substantial compressive strain. We anticipated that RP nickelate thin films might encompass mixed phases (with different $n$ values) when $\varepsilon$ spans a range from +1% to –1%. Analysis of the lattice parameters of RP nickelate thin films revealed a systematic increase in out-of-plane lattice constants as in-plane compression intensifies, indicating accommodation of substrate-induced misfit strain through structural deformation. Further, both in-plane and out-of-plane lattice constants decrease progressively with increasing compressive strain ($\varepsilon < -2.17\%$). This effect can be attributed to the formation of structural defects under substantial compressive strains, leading to the creation of various stacking faults that alleviate the misfit strains.

Microscopic analysis of RP nickelate thin films' structure was conducted using scanning transmission electron microscopy (STEM). Figures 2a and 2b show the high-angle annular dark field (HAADF) STEM images of the RP nickelate thin films grown on DSO and LAO substrates, respectively. The samples were imaged along the substrates' [110] zone axis, revealing clear and abrupt interfaces between RP nickelate films and the substrates, marked by white dashed lines. Distinct electron diffraction patterns were observed in the Fourier transform data. While RP reflections were prominent in RP nickelate films on LAO, these similar diffraction patterns were notably absent in the RP films on DSO. Detailed high-magnification

HAADF-STEM images from selected regions within the RP nickelate films are presented in Figures 2c and 2d. The atom arrangements shown in the STEM images illustrate the crystal structures of LaNiO$_3$ and La$_3$Ni$_2$O$_7$ on DSO and LAO substrates, respectively. Intensity profiles derived from line scans averaged across the respective HAADF-STEM images are shown in Figures 2e and 2f. The LaNiO$_3$ films show alternate NiO$_2$ and LaO atomic planes, maintaining approximately 3.87 Å. In contrast, for La$_3$Ni$_2$O$_7$ films, two layers of NiO$_6$ octahedra are interspersed by two LaO atomic planes, with approximately 2.89 Å. This significant contrast in atomic structures between two RP nickelate films corroborates the high structural quality observed in our XRD measurements. Moreover, HAADF-STEM images were also obtained from an RP nickelate film grown on SrTiO$_3$ (STO) substrates (Figure S3). Notably, both LaNiO$_3$ and other high-order RP nickelate phases were discernible due to the intermediate strain states. This observation underscores the significance of strain-mediated phase transition in RP nickelate thin films.

To elucidate the observed coexistence of multiple RP phases and strain-mediated phase transition, we conducted first-principles density-functional theory calculations for the formation energy and lattice structures of RP nickelates (see Method). Initially, the in-lattice constants for La$_3$Ni$_2$O$_7$, La$_4$Ni$_3$O$_{10}$, and LaNiO$_3$ were computed as, 3.866Å, 3.868Å, and 3.832Å respectively. These values not only exhibit consistency with our measured values (~3.83Å) but also reveal an intrinsic transition from $n$ = 2 to 3 under tensile strain. Additionally, by changing valence states of Ni ions using the virtual crystal approximation, we deviated the valence state of Ni in La$_3$Ni$_2$O$_7$, La$_4$Ni$_3$O$_{10}$, and LaNiO$_3$, from original states of Ni$^{2.5+}$, Ni$^{2.67+}$, and Ni$^{3+}$, respectively, within a reasonable range. This gradual change was achieved by adjusting the valence state of La from +2.5 to +3.5, beyond the original +3 state. Figures 3a and 3b shows the calculation results for LaNiO$_3$ and La$_3$Ni$_2$O$_7$, respectively. A clear trend emerges: higher valence states of Ni are favorable under larger in-plane lattice constants (tensile strain), while lower valence states are favorable for smaller in-plane lattice constants. This tendency demonstrates the smaller in-plane lattice constants typically indicate stronger $d$-$p$ hybridization, leading to more effective trapping of holes by O-$p$ orbitals, resulting in Ni ions having more electrons to occupy and consequently adopting a higher valence state. Conversely, larger in-plane lattice constants reduce *d-p* hybridization, causing the O cations to be closer to $O^{2-}$ and leading to Ni ions in lower nominal states. Another noteworthy point in Figure 2a is that compressive strain appears capable of triggering transitions from $Ni^{3+}$ to $Ni^{(3-x)+}$, where x is between 0 to 0.5. It's essential to note that this region encompasses the states of $Ni^{2.5+}$ and $Ni^{2.67+}$. This demonstrates not only a strong coupling between lattice and Ni states but also the role of increasing in-plane lattices in triggering the phase transition from layered nickelates (*n* = 2, 3) to three-dimensional nickelates (*n* =∞).

Since the results of $La_4Ni_3O_{10}$ is remarkably similar to that of $La_3Ni_2O_7$, we will mainly discuss the calculation results of $La_3Ni_2O_7$. When increasing the valence states of Ni ions from the original $Ni^{2.5+}$ to higher values, the in-plane lattice constants increase from 3.866 to 3.935Å. This behavior mirrors that observed in $LaNiO_3$, where higher valence states of Ni are favorable for larger in-plane lattices. However, when the valence states of Ni ions are further reduced to +2.4 and to +1.9, the in-plane lattice appears to be situated in a frozen region spanning the lattice of $La_3Ni_2O_7$ between 3.866 and 3.850Å. This lattice fluctuation seems to originate from the complex bands and their splitting of *n* = 1 to 3, strongly suggesting the instability of $La_3Ni_2O_7$ under compressive strain and an underlying driving force within multiple RP phases under compressive strain. Based on the above findings, we attribute the transition from *n* = 2 to *n* > 2 and *n* = ∞ to a dual driving force resulting from both larger in-plane lattices and charge instability of $Ni^{2.5+}$ state compared with $Ni^{2.67+}$ and $Ni^{3+}$ states under tensile strain. Meanwhile, the transition from *n* = 2 to *n* > 2 under compressive strain conditions appears to originate solely from the charge instability of $Ni^{2.5+}$ under compressive strain, this has similarities to the structure co-existence in Sr-doped nickelates [38].

Misfit strain not only induces structural deformation in RP nickelate thin films but also alters their transport behavior accordingly. Figure 4a shows the temperature-dependent resistivity (ρ) curves of RP nickelate films under various strain states. When ε = 3.44% (on GSO) and ε = -4.12% (on YAO), the RP nickelate films exhibit insulating behavior across all temperatures, despite having relatively low resistivities within the range of $10^{-2}$–$10^{-3}$ Ω·cm. This effect may be attributed to the formation of structural dislocations and oxygen vacancies within the films under substantial misfit strains. Within the moderate strain regime, the RP nickelate films demonstrate a clear metal-to-insulator transition (MIT) as the temperature decreases. MIT behavior can be attributed to the potential density-wave like transition observed in bulk RP nickelates and predicted by other theorists [39-42], or simply due to the mixture of multiple phases with different electronic ground states (will discuss later). Figures 4b and 4c illustrate the metal-to-insulator transition temperature ($T_{MI}$) and room-temperature resistivity ($\rho_{300K}$) as functions of ε, respectively. The RP nickelate films exhibit minimum $\rho_{300K}$ and $T_{MI}$ under the smallest tensile strain (ε = 0.97%), suggesting an extremely sensitive dependence of transport behavior on misfit strain.

Moreover, optical ellipsometer measurements were conducted on the RP nickelate thin films to comprehend strain-dependent optical constants and its relations with photon excitations. Figures 4d and 4e show the optical conductivity [$\sigma_1(\omega) = \omega\varepsilon_2(\omega)/4\pi$] spectra as a function of photon energy ($h\omega$), where $\varepsilon_2(\omega)$ represents the imaginary part of dielectric constants (Figure S4). Across all films, a suppressed Drude feature near 0 eV was observed, consistent with relatively high electrical resistivity at room temperature. As the tensile strain decreases, RP nickelate films transition from $LaNiO_3$ to $La_3Ni_2O_7$, resulting in additional electron filling in the $e_g$ orbital. This leads to the closure of the optical gap below a photon energy of 1 eV (Figure 4d). Conversely, with increasing compressive strain, defects such as oxygen vacancies and dislocations cause a greater band population favoring the $A^1$ transition. Consequently, spectroscopic features in $\sigma_1(\omega)$ shift significantly toward lower photon energies, closing the optical band gap. The $A^2$ transition, positioned at ~1.45 eV, is attributed to the occupied Ni $e_g$ to unoccupied $e_g$ transition. In the intermediate strain region with mixed-phase RP nickelates (ε from -2.17% to 1.88%), splintered $A^2$ transitions occur due to the coexistence of $Ni^{3+}$ and $Ni^{2+}$ ions. The $B^1$ optical absorptions exhibit a substantial shift from 3 to 4 eV with decreasing tensile strain. The engagement of the filling low-Hubbard (LH) band with the reduction in $B^2$ optical transition involves a transition from Ni $t_{2g}$ to LH band ($e_g$). These optical conductivity results on strained RP nickelates provide a comprehensive understanding of photon excitations from filled bands to unoccupied states, in consistent with our transport measurements.

To further intricately correlated the transport behavior with their intrinsic electronic configurations, we conducted elemental-specific X-ray absorption spectroscopy (XAS) measurements on the RP nickelate films under three distinct strain states ($\varepsilon$ = -1.12%, 1.88%, and 3.44%) at room temperature. Figure 5a illustrates the XAS results at O $K$-edges for the three representative RP nickelate thin films. The red shaded region, centered at ~527.5 eV, distinctly displays a characteristic low-energy pre-peak stemming from the ligand holes [43,44]. Notably, when compared to reference samples (LaNiO$_3$ and NiO bulks) [45,46], the pre-peak of compressively strained RP nickelate shifts to higher energy, suggesting a change in the Ni $d$-band configuration due to filled oxygen ligand holes with electrons. Within the blue shaded region, where center position is around ~532 eV, we observe the excitation of electrons from O 1$s$ to the Ni 3$d$-O 2$p$ hybridization states. The peak intensity at ~532 eV gradually increases as misfit strain decreases from tensile to compressive, indicating enhanced hybridization between Ni$^{2+}$ and O$^{2-}$ ions in the compressively strained RP nickelates. Consequently, compressive strain induces the formation of a $d^8$ state and suppresses the $d^7$ configuration in RP nickelate films. With the profoundly modified $d$-band filling on Ni ions established, we delved into the orbital occupations using X-ray linear dichroism (XLD). Linearly polarized X-rays were incident on the sample surface at angles of 30° and 90° with respect to the surface plane. Figures 5b and 5c exhibit XAS and XLD at Ni $L_2$-edges for RP nickelates under different strain states, respectively. In tensile-strained RP nickelate films, the slight discrepancy between the peak energies of I$_{90°}$ and I$_{30°}$ suggests a preference for electrons to occupy $d_{x^2-y^2}$ orbitals, resulting in a negative XLD value. Conversely, compressively strained RP nickelate films exhibit positive XLD signals, indicating a higher affinity for electrons to occupy $d_{3z^2-r^2}$ orbitals. A notable aspect of XLD data for compressively strained RP nickelate films is the presence of $e_g$ band splitting, which is absent in the tensile-strained RP nickelate films. Direct inspection of the energy positions for in-plane (~870.4 eV) and out-of-plane (~870.6 eV) absorption curves at XAS Ni $L_2$-edges reveals an energy difference of ~ 0.2 eV, indicating $e_g$ band splitting between states with Ni $d_{x^2-y^2}$ and $d_{3z^2-r^2}$ orbital characters (inset of Figure 5b). This observation lends support to the idea that band splitting and orbital polarization in RP nickelates arise from misfit-induced structural distortion, agreeing well with recent theoretical calculations [10,17].

Finally, we demonstrate the ability to switch the formation of different RP-phase nickelate films by altering the oxygen partial pressure. Under identical misfit strain conditions (on LAO), we observed that the RP nickelate films predominantly favor the $La_3Ni_2O_7$ dominated phase (D-$La_3Ni_2O_7$) when $P_{O2}$ = 0.25 Torr. Further increasing $P_{O2}$ will lead to the degraded sample quality, possibly introducing more structural defects and non-stoichiometry in the ratio between cations. Conversely, when $P_{O2}$ decreases to 0.1-0.15 Torr, the films are dominated by the $La_4Ni_3O_{10}$ phase (D-$La_4Ni_3O_{10}$). A pure highly insulating $La_2NiO_4$ phase will form when $P_{O2}$ below 0.05 Torr (Figure S5). In Figure S6a, XRD θ-2θ curves from D-$La_3Ni_2O_7$ and D-$La_4Ni_3O_{10}$ thin films were directly compared to the reference data of pure $La_3Ni_2O_7$ and $La_4Ni_3O_{10}$ (shown as dashed lines) [47]. Both films exhibit distinct diffraction patterns, highlighting the sensitivity of phase formation to the oxygen environment during film growth. STEM measurements on both D-$La_3Ni_2O_7$ and D-$La_4Ni_3O_{10}$ were conducted to analyze the relative concentrations of different phases in the RP nickelate films. Figures S6b and S6c display representative HAADF- and annular-bright-field (ABF)-STEM images of D-$La_4Ni_3O_{10}$ thin films, revealing one $La_3Ni_2O_7$ layer sandwiched between two $La_4Ni_3O_{10}$ layers. Both D-$La_3Ni_2O_7$ and D-$La_4Ni_3O_{10}$ show minimal octahedral tilt, indicating Ni-O-Ni bond angles close to 180° at ambient conditions. Figures S6d present estimated concentrations of $La_3Ni_2O_7$ and $La_4Ni_3O_{10}$ phases in RP nickelate films. Despite XRD results suggesting a "*pure*" RP phase macroscopically in both films, the microscopic views demonstrate approximately ~18% $La_4Ni_3O_{10}$ in D-$La_3Ni_2O_7$ and ~27% $La_3Ni_2O_7$ in D-$La_4Ni_3O_{10}$. We conducted ρ-T curve measurements for both D-$La_3Ni_2O_7$ and D-$La_4Ni_3O_{10}$, as shown in Figure S6e. Pure $La_3Ni_2O_7$ appears highly insulating, while $La_4Ni_3O_{10}$ exhibits a metallic phase. The resistivities of mix-phase RP nickelate films lie between those of pure phases, in consistent with our structural analysis. The similar phase formation energies among different competing RP phases make fabricating pure single-phase RP nickelate films extremely challenging, not only in thin film but also in the single crystals. It is crucial to thoroughly examine discussions concerning transport behavior in mix-phase RP nickelates, particularly regarding high-temperature superconductivity in previously reported "*pure*" single-crystalline $La_3Ni_2O_7$. The coexistence of an insulating $La_3Ni_2O_7$ and a metallic high-order RP nickelate may explain the low-temperature resistivity upturn at ambient pressure. Identifying the precise superconducting phase under high-pressure conditions is crucial for advancing discussions and theoretical models that elucidate the observed HT superconductivity in RP nickelates [15]. Our thin film experiments highlight a significant aspect: mixed-phase nickelates exhibit additional grain boundaries, which occupy only a minuscule fraction of the sample volume. It is imperative to conduct a comprehensive investigation into the structural characteristics and their correlations with transport properties. This examination is pivotal in comprehending the origins of pressure-induced HT superconductivity and the absence of strong diamagnetic signals in RP nickelates.

**Summary**

In summary, our study unveils the strain-dependent structural and transport properties of RP phase nickelate thin films. Tensile strain stabilizes high-order RP nickelates, whereas compressive strain favors $La_3Ni_2O_7$ thin films. Systematic optical conductivity and X-ray absorption spectroscopy measurements delineate the evolution of band structure and electronic states in RP nickelate thin films as a function of misfit strain. The observed band splitting energy (~ 0.2 eV) in compressively strained RP nickelate films aligns with recent theoretical calculations. It's worth highlighting that as-grown RP nickelates invariably contain mixed RP phases, posing challenges for confirmation through macroscopic structural characterizations like X-ray or neutron diffractions. Particular attention is crucial when discussing phenomena such as the metal-to-insulator transition at low temperatures and pressure-induced superconductivity. We anticipate that biaxially compressively strained pure-phase $La_3Ni_2O_7$ or other high-order RP nickelate thin films could potentially reduce the superconducting transition pressure—a straightforward avenue for future research. Our work serves as a foundational step, delineating a clear path for future studies on single-phase RP nickelate films using comprehensive investigative tools and theoretical calculations.

**Experimental and calculation details**

<u>Thin film synthesis and structural characterizations</u>

For the PLD growth process, we prepared a stoichiometric $La_3Ni_2O_7$ ceramic target through a solid reaction route using a precise chemical ratio from a mix of $La_2O_3$ and NiO powder. X-ray diffraction measurements on the target demonstrate a stoichiometric and correct chemical composition. RP nickelates thin films with thickness approximately 30 nm were fabricated on various substrates with different lattice parameters. To investigate the impact of strain on the physical properties, all films were deposited under identical experimental conditions ($T_{sub}$ = 750 °C, $P_{O2}$ = 0.3 Torr, and $E_{laser}$ = 1.25 J/cm$^2$). After the growth, the films were post-annealed under high oxygen partial pressure at $P_{O2}$ = 100 Torr and then cooled down slowly to room temperature. This process typically can largely remove oxygen vacancies in order to maintain their stoichiometry. Crystallographic analysis, that is, X-ray diffraction 2θ-ω, X-ray reflectivity (XRR), and reciprocal space mapping (RSM), were carried out using a Panalytical X'Pert3 MRD diffractometer with Cu Kα1 radiation equipped with a 3D pixel detector. Cross-sectional TEM specimens with different crystallographic and strain states were prepared using Ga$^+$ ion milling after the mechanical thinning. The HAADF and ABF imaging were performed in the scanning mode using JEM ARM 200CF microscopy at the Institute of Physics (IOP), Chinese Academy of Sciences (CAS).

Electronic state measurements

Elemental specific XAS measurements were performed on RP nickelates at the beamline 4B9B of Beijing Synchrotron Radiation Facility (BSRF). All spectra were collected at room temperature in total electron yield (TEY) mode. Our samples are properly electrically connected to the sample holder using silver glue. The XLD measurements were performed by changing the incident angle of the linearly polarized X-ray beam. The X-ray scattering plane was rotated by 90° and 30° with respect to the sample surface plane. The XAS signals are normalized to the values at the pre- and post-edges. When the X-ray beam is perpendicular to the surface plane (90°), XAS signal directly reflects the $d_{x^2-y^2}$ orbital occupancy. While the angle between the X-ray beam and surface plane is 30°, the XAS signal contains orbital information from both $d_{x^2-y^2}$ and $d_{3z^2-r^2}$ orbitals. For simplifying the discussions, the XLD signals were calculated by $I_{90°} - I_{30°}$. The XLD signal directly reflects the orbital polarization of a sample under different strain states. The optical conductivity of different strained RP nickelate thin films were measured at room temperature using a commercial optical ellipsometer (J. A. Woollam Co., Inc.).

Density-functional theory simulations

We utilized Density Functional Theory (DFT) as the fundamental computational framework to explore the electronic and structural properties of RP phase nickelates. The Vienna Ab initio Simulation Package (VASP) served as our computational tool. The Perdew-Burke-Ernzerhof (PBE) version of the generalized gradient approximation (GGA) exchange-correlation functional was selected to accurately describe electronic interactions. We expanded electronic wave-functions using a plane-wave basis set, setting a kinetic energy cut-off of 500 eV to ensure the convergence of total energy calculations. Our systematic benchmarks and tests affirm that this cut-off energy strikes an optimal balance between computational efficiency and result accuracy, establishing a robust foundation for our investigations. The k-mesh for the $LaNiO_3$, $La_3Ni_2O_7$ and $La_4Ni_3O_{10}$ is 15×15×15, 9×9×3, and 9×9×3, respectively. To modify the valence states of Ni ions, we employed the virtual crystal approximation (VCA) implemented in VASP. In this approach, divalent and tetravalent elements were chosen to occupy the nearest positions to La in the element table, specifically Ba and Ce. This tailored adjustment in valence states enhances the accuracy of our simulations, providing a more representative description of the electronic structure in RP nickelates.

## Acknowledgements

We appreciate the thoughtful discussions from Prof. Y. F. Nie (Nanjing University), Prof. L. F. Wang (University of Science and Technology in China), Prof. Wei Li (Institute of Theoretical Physics, CAS), Prof. Yanwei Cao (Ningbo Institute of Science and Technology, CAS), and Prof. Kun Jiang (Institute of Physics, Chinese Academy of Sciences). This work was supported by the National Key Basic Research Program of China (Grant Nos. 2020YFA0309100 and 2019YFA0308500), the National Natural Science Foundation of China (Grant Nos. 11974390, U22A20263, 52250308), the CAS Project for Young Scientists in Basic Research (Grant No. YSBR-084), the China Postdoctoral Science Foundation (Grant No. 2022M723353), the Special Research assistant of Chinese Academy of Sciences, the Guangdong-Hong Kong-Macao Joint Laboratory for Neutron Scattering Science and Technology, and the Strategic Priority Research Program (B) of the Chinese Academy of Sciences (Grant No. XDB33030200). Synchrotron-based XAS and XLD experiments were performed at the beamline 4B9B of the Beijing Synchrotron Radiation Facility (BSRF) via user proposals.

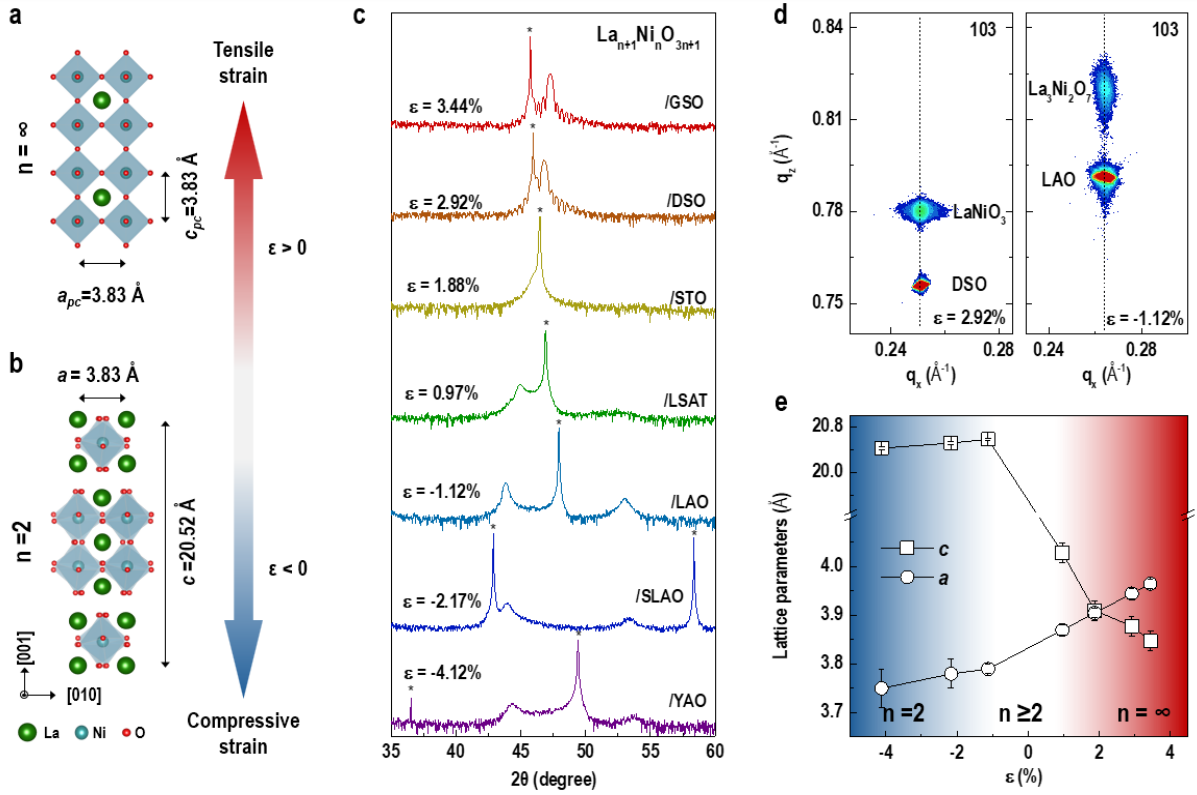

**Figure 1. Structural evolution of RP nickelate thin films under strain**. (**a**) and (**b**) Atomic structures of LaNiO$_3$ (RP, $n = \infty$) and La$_3$Ni$_2$O$_7$ (RP, $n = 2$), respectively. Under tensile strain ($\varepsilon > 0$), the as-grown RP nickelate films have $n$ close to $\infty$, whereas the RP nickelate films favor $n = 2$ under compressive strain ($\varepsilon < 0$). (**c**) XRD $\theta$-$2\theta$ scans for RP nickelate films around substrates' peaks (indicated with *) under different strain states. (**d**) RSMs of RP nickelate films around DSO and LAO substrates' 103 diffraction peaks, respectively. (**e**) Lattice parameters of RP nickelate films as a function of misfit strain.

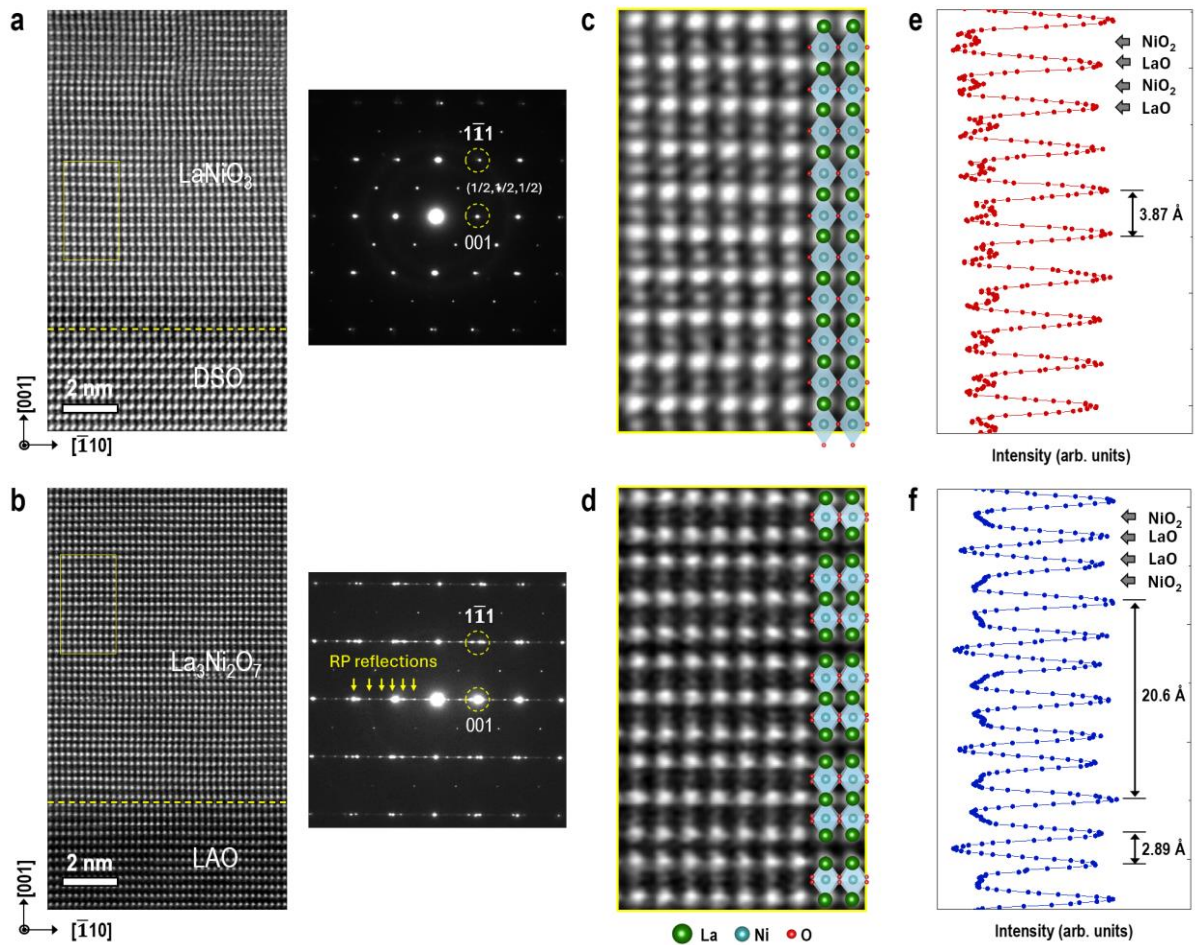

**Figure 2. Microscopic views of RP nickelate films under strain**. (**a**) and (**b**) HAADF-STEM images of RP nickelate films grown on DSO and LAO substrates, respectively. SEAD patterns from film regions were shown on the right side of each STEM images. (**c**) and (**d**) High-magnification RP nickelate films with distinct structures, respectively. The atomic structures were illustrated as guide for eyes. (**e**) and (**f**) Intensity profiles obtained from the line scans averaged across the respective HAADF-STEM images of (**c**) and (**d**), respectively.

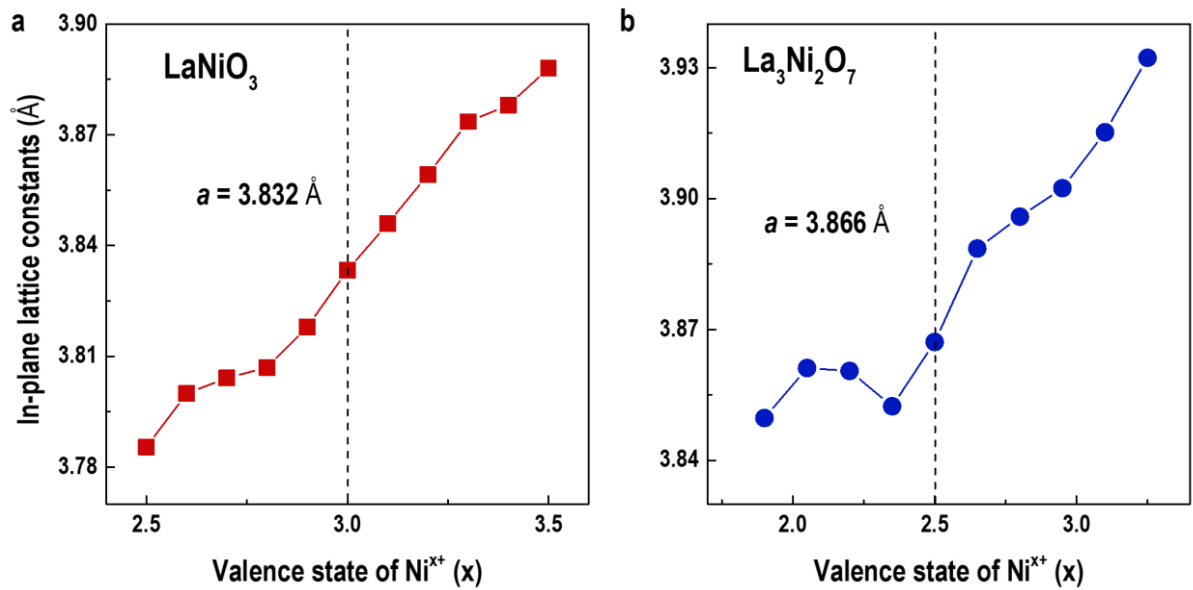

**Figure 3. Correlation between in-plane lattices and the valence state of Ni ions** in (a) LaNiO$_3$ and (b) La$_3$Ni$_2$O$_7$, where the vertical lines in (a) and (b) indicate the nominal valence state of Ni ions in LaNiO$_3$ (Ni$^{3+}$) and La$_3$Ni$_2$O$_7$ (Ni$^{2.5+}$), respectively.

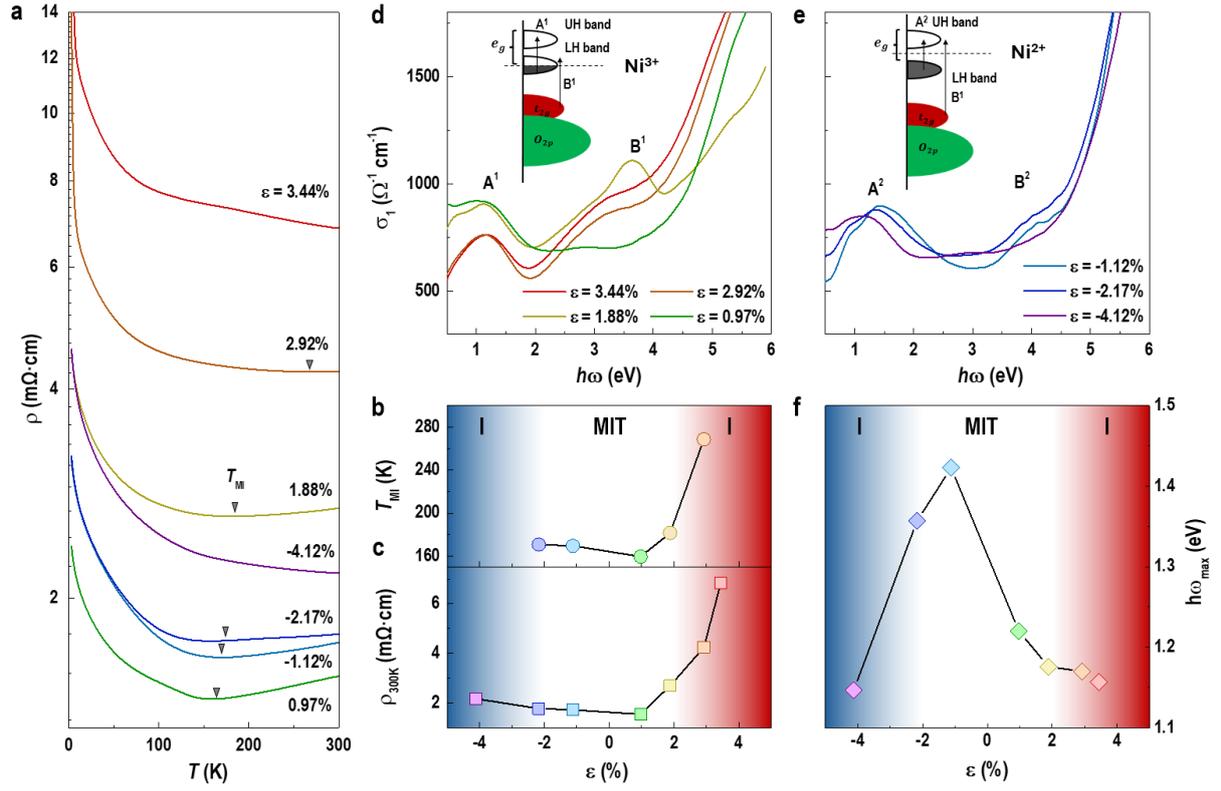

**Figure 4. Transport behavior and optical conductivity of strained RP nickelate films**. (**a**) ρ-T curves of strained RP nickelate films. The resistivity minimum of ρ-T curves was marked as "▼". (**b**) Metal-to-insulator transition temperature ($T_{MI}$) and (**c**) room temperature resistivity ($\rho_{300K}$) as a function of misfit strain. Optical conductivity of (**d**) tensile strained and (**e**) compressively strained RP nickelate films, obtained from spectroscopic ellipsometry at room temperature. Insets show the schematics of band structures. (**f**) The peak position ($h\omega_{max}$) at the optical conductivity maximum as a function of misfit strain.

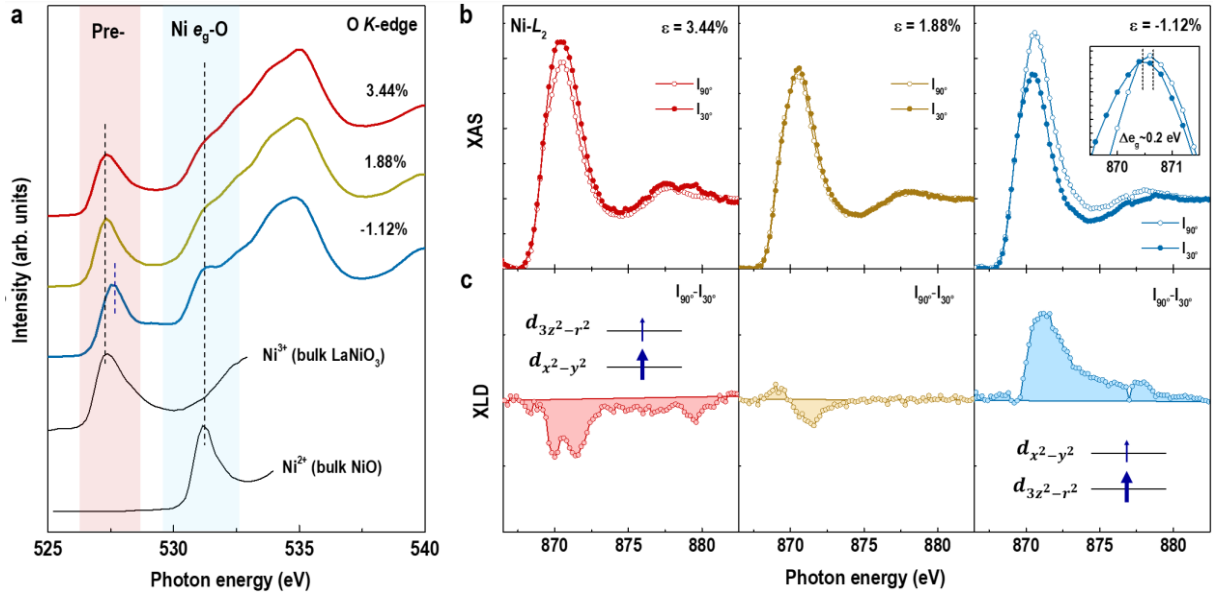

**Figure 5. XAS of strained RP nickelate films**. (a) XAS at O $K$-edges for strained RP nickelate films. Dashed lines show the approximate peak positions for two main features: the pre-peak (~ 527.5 eV) and hybridized Ni $e_g$-O (~ 532 eV) states by comparing with two spectra of reference samples bulk LaNi$^{3+}$O$_3$ and bulk Ni$^{2+}$O. (b) XAS and (c) XLD at Ni $L_2$-edge for strained RP nickelate films, respectively. Linearly polarized X-ray were used to measure XAS of strained RP nickelate films with incident angle of 30° and 90°. XLD were estimated from intensity difference between I$_{30°}$ and I$_{90°}$.

Supplementary Materials for

# Strain-mediated phase crossover in Ruddlesden−Popper nickelates

Ting Cui,[1,2] Songhee Choi,[1] Ting Lin,[1,2] Chen Liu,[3] Gang Wang,[1,2] Ningning Wang,[1,2] Shengru Chen,[1,2] Haitao Hong,[1,2] Dongke Rong,[1,2] Qianying Wang,[1,2] Qiao Jin,[1] Jia-Ou Wang,[3] Lin Gu,[4] Chen Ge,[1] Can Wang,[1,2,5] Jin Guang Cheng,[1] Qinghua Zhang,[1,2,5] Liang Si,[6,7,*] Kui-juan Jin,[1,2,5,*] and Er-Jia Guo[1,2,*]

[1] Beijing National Laboratory for Condensed Matter Physics and Institute of Physics, Chinese Academy of Sciences, Beijing 100190, China

[2] Department of Physics & Center of Materials Science and Optoelectronics Engineering, University of Chinese Academy of Sciences, Beijing 100049, China

[3] Institute of High Energy Physics, Chinese Academy of Sciences, Beijing 100049, China

[4] National Center for Electron Microscopy in Beijing and School of Materials Science and Engineering, Tsinghua University, Beijing 100084, China

[5] Songshan Lake Materials Laboratory, Dongguan, Guangdong 523808, China

[6] School of Physics, Northwest University, Xi'an 710127, China

[7] Institut für Festkörperphysik, TU Wien, 1040 Vienna, Austria

*Correspondence and requests for materials should be addressed to L.S., K.J.J. and E.J.G. (emails: siliang@nwu.edu.cn, kjjin@iphy.ac.cn and ejguo@iphy.ac.cn)

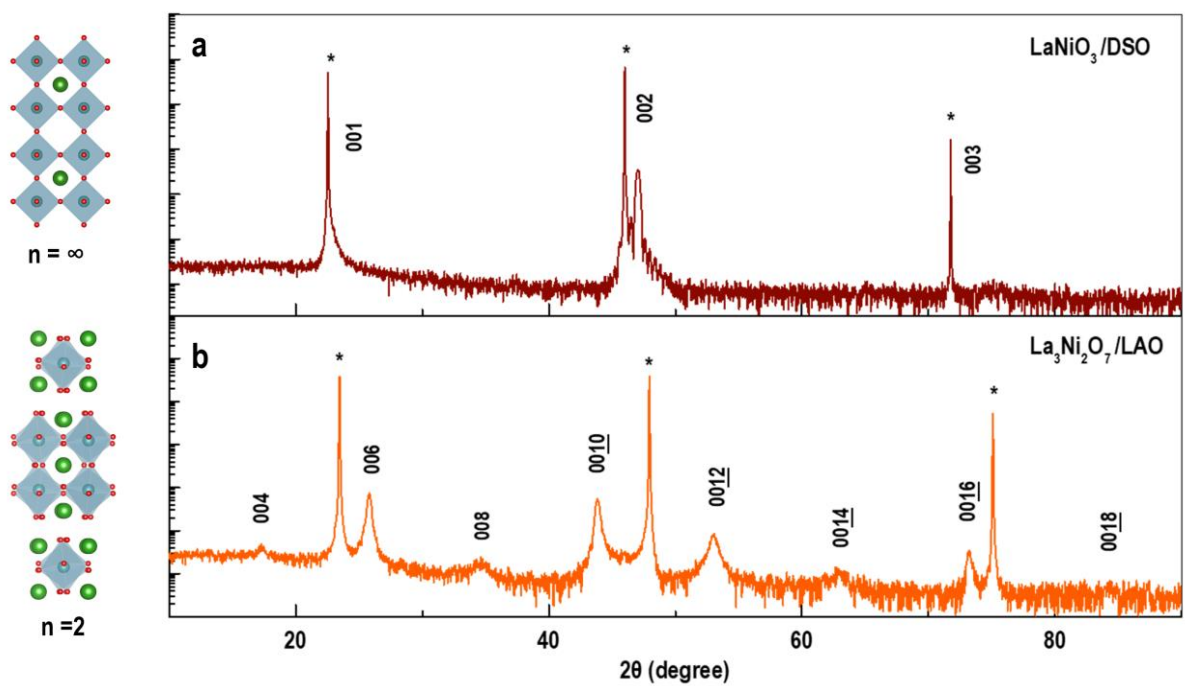

**Figure S1. Wide-range XRD θ-2θ scans** for LaNiO$_3$ and La$_3$Ni$_2$O$_7$ films grown on DSO and LAO substrates (peaks marked with *), respectively. The atomic structures of LaNiO$_3$ and La$_3$Ni$_2$O$_7$ are present on the left side of XRD curves.

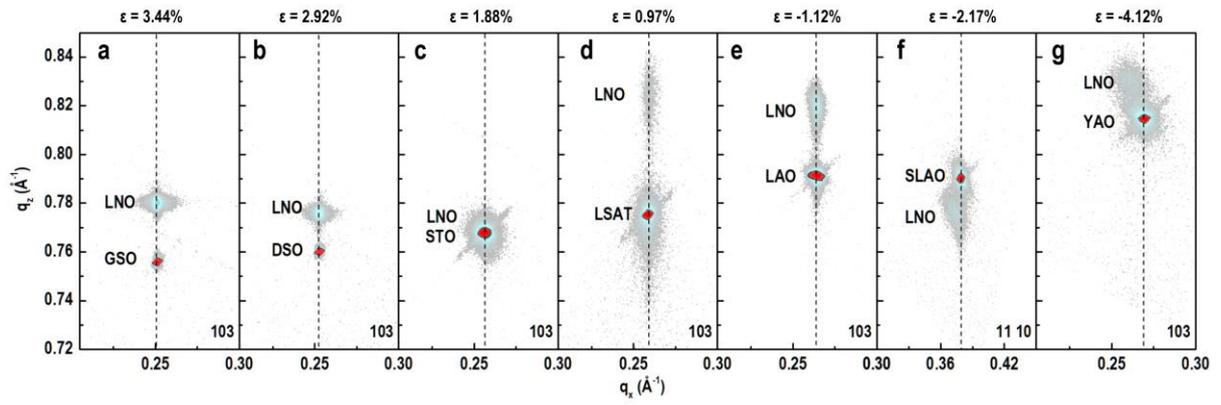

**Figure S2. RSM of RP nickelate films on various substrates.** Except for large compressive strain (on SLAO and YAO substrates), all RP nickelate films are coherently strained.

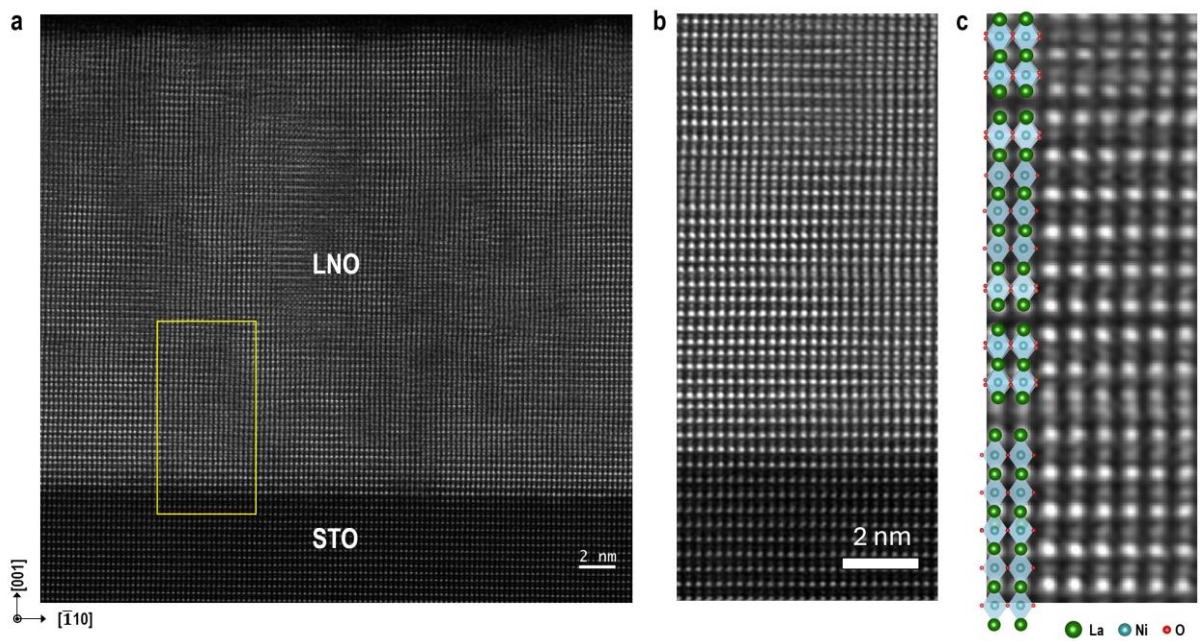

**Figure S3. STEM image of RP nickelate films grown on STO substrates**. (a) Low-magnification HAADF-STEM image. (b) Interface region. (c) Zoom-in microstructure of RP nickelates. The RP nickelate films present a mixed phase composed of $LaNiO_3$ and $La_3Ni_2O_7$ when it is grown on STO substrates. Obviously, we can observe the stacking defaults at the grain boundaries.

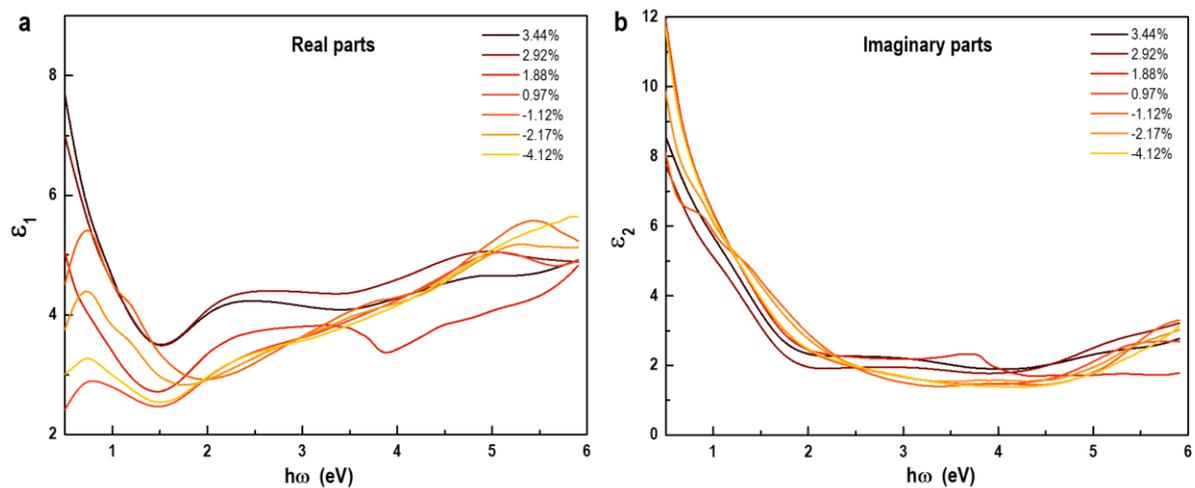

**Figure S4. Experimental ε₁ and ε₂ of strained RP nickelate films.** The ellipsometry measurements were carried out at room temperature. These data are used to calculate the optical conductivities $\sigma_1(\omega)$.

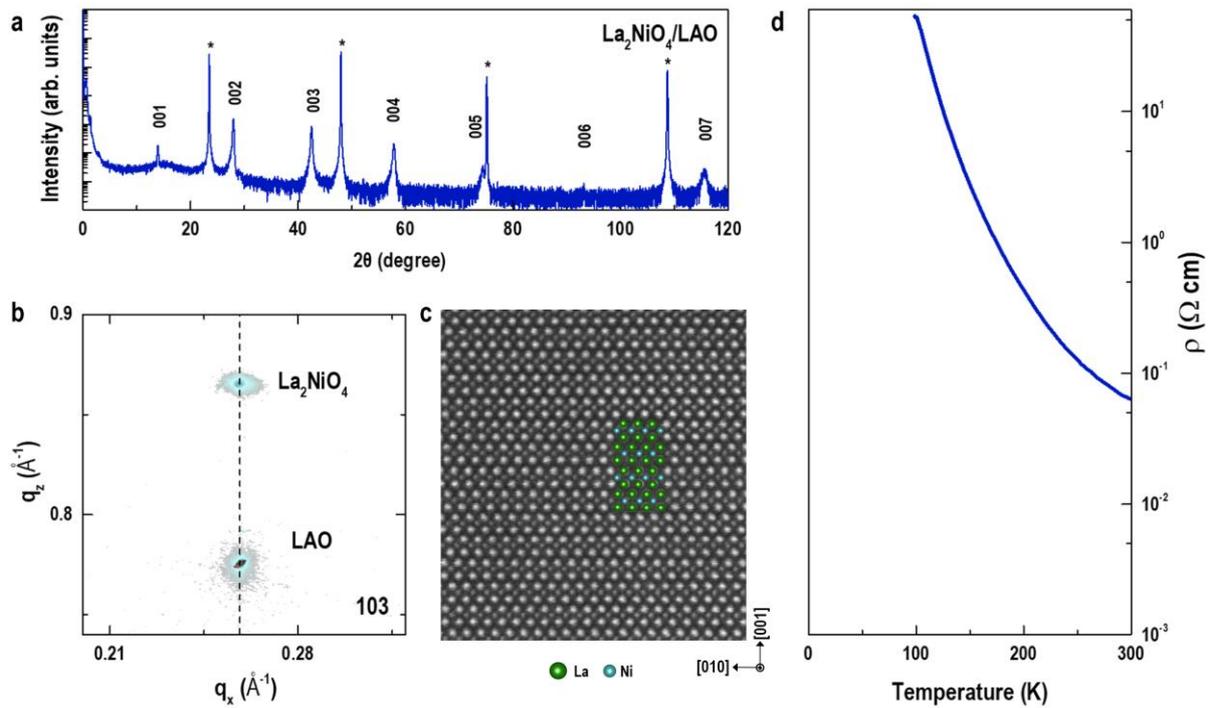

**Figure S5. Structural and transport behavior of La₂NiO₄ thin films**. (**a**) XRD θ-2θ scan and (**b**) RSM of a La$_2$NiO$_4$ film grown on LAO substrate (peaks marked with *). (**c**) HAADF-STEM image of a La$_2$NiO$_4$ film. (**d**) ρ-T curves of a La$_2$NiO$_4$ film, indicating an insulating nature at low temperatures.

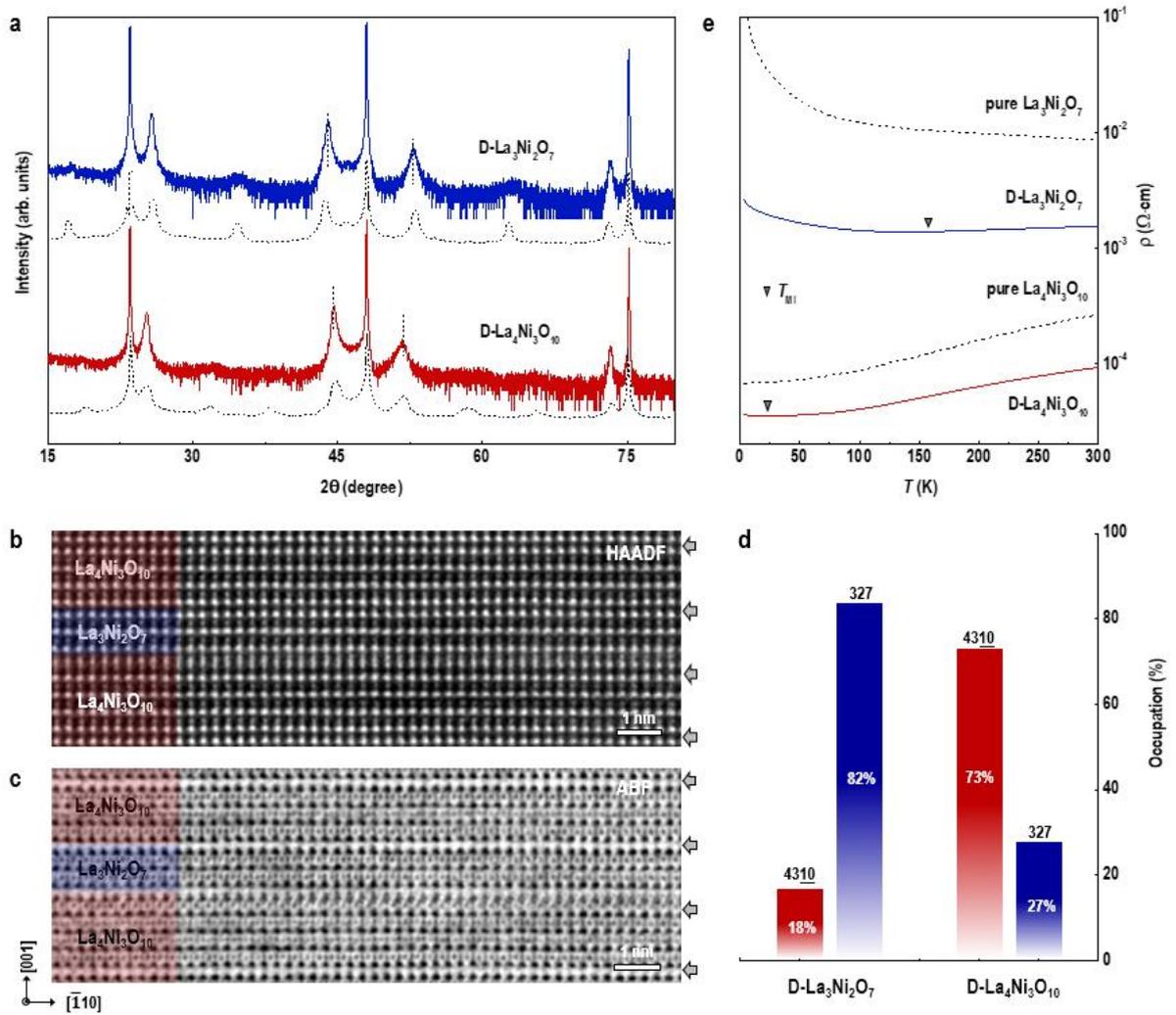

**Figure S6. Mixed RP-phase nickelate films.** (**a**) Wide-range XRD $\theta$-$2\theta$ scans for RP nickelate films with $La_3Ni_2O_7$ dominated (D-$La_3Ni_2O_7$) and $La_4Ni_3O_{10}$ dominated (D-$La_4Ni_3O_{10}$) phases. Dashed lines show the reference XRD curves for pure $La_3Ni_2O_7$ and $La_4Ni_3O_{10}$. (**b**) HAADF- and (**c**) ABF-STEM images of D-$La_4Ni_3O_{10}$ films, respectively. Black arrows indicate the positions of LaO-LaO interlayer. (**d**) Occupation of different RP phases within D-$La_3Ni_2O_7$ and D-$La_4Ni_3O_{10}$ films. (**e**) $\rho$-T curves of D-$La_3Ni_2O_7$ and D-$La_4Ni_3O_{10}$ films. Dashed lines show the reference $\rho$-T curves of pure $La_3Ni_2O_7$ and $La_4Ni_3O_{10}$.